\documentclass[letterpaper]{article} 
\usepackage{aaai25}  
\usepackage{times}  
\usepackage{helvet}  
\usepackage{courier}  
\usepackage[hyphens]{url}  
\usepackage{graphicx} 
\usepackage{natbib}  
\usepackage{caption} 
\usepackage{amsmath,amsfonts}
\usepackage{booktabs,multirow,rotating,array}
\usepackage{makecell}
\usepackage{xcolor}
\usepackage{bbding}
\usepackage{utfsym}
\usepackage{natbib,tabularx}
\usepackage[misc]{ifsym} 

\usepackage[ruled,vlined]{algorithm2e}
\usepackage{newfloat}
\usepackage{listings}
\DeclareCaptionStyle{ruled}{labelfont=normalfont,labelsep=colon,strut=off} 
\title{3DMambaIPF: A State Space Model for Iterative Point Cloud Filtering via Differentiable Rendering}
\author{
    Qingyuan Zhou\textsuperscript{\rm 1}, Weidong Yang\textsuperscript{\rm 1}\thanks{Corresponding authors}, Ben Fei\textsuperscript{\rm 2}$^*$, Jingyi Xu\textsuperscript{\rm 1}, Rui Zhang\textsuperscript{\rm 1}, Keyi Liu\textsuperscript{\rm 1}, Yeqi Luo\textsuperscript{\rm 1},\\ Ying He\textsuperscript{\rm 3}
}

\affiliations{
    \textsuperscript{\rm 1}School of Computer Science, Fudan University \\ \textsuperscript{\rm 2}Department of Information Engineering, The Chinese University of Hong Kong \\\textsuperscript{\rm 3}College of Computing and Data Science, Nanyang Technological University\\
    \{zhouqy23, jyxu22, ruizhang22, 23210240242, 23212010018\}@m.fudan.edu.cn; wdyang@fudan.edu.cn; benfei@cuhk.edu.hk; yhe@ntu.edu.sg
}

\usepackage{bibentry}

\begin{document}

\maketitle

\begin{abstract}
Noise is an inevitable aspect of point cloud acquisition, necessitating filtering as a fundamental task within the realm of 3D vision. 
Existing learning-based filtering methods have shown promising capabilities on commonly used datasets. 
Nonetheless, the effectiveness of these methods is constrained when dealing with a substantial quantity of point clouds. 
This limitation primarily stems from their limited denoising capabilities for dense and large-scale point clouds and their inclination to generate noisy outliers after denoising.
To deal with this challenge, we introduce \textbf{3DMambaIPF}, for the first time, exploiting Selective State Space Models (SSMs) architecture to handle highly-dense and large-scale point clouds, capitalizing on its strengths in selective input processing and large context modeling capabilities.
Additionally, we present a robust and fast differentiable rendering loss to constrain the noisy points around the surface. 
In contrast to previous methodologies, this differentiable rendering loss enhances the visual realism of denoised geometric structures and aligns point cloud boundaries more closely with those observed in real-world objects.
Extensive evaluations on commonly used datasets (typically with up to 50K points) demonstrate that 3DMambaIPF achieves state-of-the-art results. Moreover, we showcase the superior scalability and efficiency of 3DMambaIPF on highly dense and large-scale point clouds with up to 500K points compared to off-the-shelf methods.
\begin{links}
\link{Code}{https://github.com/TsingyuanChou/3DMambaIPF}.
\end{links}
\end{abstract}

%

\section{Introduction}\label{sec:introduction}

Point clouds, assemblies of numerous three-dimensional points denoting spatial positions, play a crucial role in representing scenes across multimedia and computer vision (CV) domains.
Nevertheless, as point clouds from real-world environments are captured by devices such as LiDAR, depth cameras, or 3D scanners, noises are inevitably introduced during the scanning process~\cite{fei2024progressive}.
The presence of these noises undermines the required data consistency and accuracy in the application scenarios of point clouds, making filtering of point clouds a fundamental task within the realm of 3D vision~\cite{fei2022comprehensive,fei2023self}.
Additionally, due to the potential large-scale nature of these real-world scans, point clouds can exhibit substantial coordinate ranges and be highly dense, containing a significant number of points.


Traditional point cloud filtering methods rely on expert priors, but they are limited by the accuracy of these priors and are not suitable for complex and sparse point clouds.
In recent years, deep learning-based point cloud filtering methods have emerged prominently. 
These methods commonly employ iterative modules that consist of stacked encoder-decoder pairs. Such architectures leverage the output of the previous module as the input for the subsequent module, thereby demonstrating superior performance.
For instance, IterativePFN~\cite{de2023iterativepfn} employs iterative modules and adaptive loss functions to iteratively and progressively refine noisy points toward adaptive ground truth (GT), yielding promising results on both synthetic and real-world datasets.

However, these point cloud filtering methods are only applicable to small-scale point clouds and not suitable for filtering highly dense and large-scale point clouds.
Furthermore, previous works only have demonstrated satisfactory performance solely in low-noise environments, while limited to handling details in high-noise environments. 
This limitation is particularly evident in the generation of unrealistic boundaries, as well as a lack of appropriate constraints.
Consequently, the main challenges are twofold: (i) denoising for highly dense and large-scale point clouds with complex geometries, (ii) denoising for high-noise environments.

Recently, State Space Models (SSMs), particularly structured SSMs (S4)~\cite{gu2021efficiently}, have shown great promise in Natural Language Processing (NLP)~\cite{lu2024structured}. Built upon the S4 framework, Mamba~\cite{gu2023mamba} exhibits outstanding performance across many tasks in NLP and 2D CV ~\cite{he2024densemamba, yang2024clinicalmamba,zhao2024cobra,misra2024low}.
The remarkable achievements of Mamba are a testament to its linear computational complexity and exceptional long-range context learning capabilities on extensive sequences. Moreover, the integration of time-varying parameters and a hardware-aware algorithm within Mamba marks a significant leap forward in the field of sequence modeling.

However, exploiting SSM models for point cloud filtering has yet to be fully explored.
Direct application of Mamba may not fully address the challenges involved.
Therefore, we propose \textbf{3DMambaIPF}, a novel \textbf{I}terative \textbf{P}oint Clouds \textbf{F}iltering model utilizing the \textbf{Mamba} module with differentiable point rendering techniques.
Specifically, 3DMambaIPF is composed of multiple iterations of Mamba-Denoising Modules, each comprising a pair of Mamba encoders and decoders.
Initially, the input point cloud is divided into patches, within which a graph structure is generated based on points and their neighbors to extract position features.
In the Mamba-based Encoder, the Dynamic EdgeConv module encodes positional features with MLPs, which are then fed into a Mamba module to select and generate position-dependent features from sequential inputs.
The Mamba-based Decoder is responsible for upsampling features into point cloud patches.
During the training process, a novel rendering loss is introduced leveraging a differential point rendering technique, and the loss is computed by the disparities of rendered images of filtered point cloud and GT.
Subsequently, an adaptive GT altering throughout the iterations is introduced, gradually transitioning from an initially noise-added state to the true GT.
Following the completion of all iterations, a patch-stitching method is employed to reconstruct the filtered point cloud from the patches.
Experimentally, 3DMambaIPF is trained using PU-Net~\cite{yu2018pu} dataset and achieves state-of-the-art results compared to off-the-shelf methods.
3DMabaIPF is also able to well generalize to high-noise point clouds from the Stanford 3D Scanning Repository~\cite{curless1996volumetric,krishnamurthy1996fitting} for synthetic objects with complex geometries.
Moreover, 3DMambaIPF outperforms baseline methods in terms of visual perceptions on real-world scene datasets. 
In summary, our main contributions are as follows:
\begin{itemize}
\item A novel iterative Mamba-based backbone is designed for point cloud filtering, allowing for the effective modeling of long-sequence point cloud features. This approach enhances the accuracy and speed of highly dense and large-scale point cloud processing.
\item A well-designed differentiable point rendering loss is introduced into 3DMambaIPF, supplanting the distance-based loss functions. This differentiable rendering loss endows 3DMambaIPF with the ability to handle edge denoising and makes the denoising results more realistic.
\item 3DMambaIPF surpasses existing methods by achieving state-of-the-art results on the commonly used PU-Net dataset, while demonstrating excellent performance even on highly-dense and large-scale datasets with high noise added.
\end{itemize}

\section{Related Work}
\begin{figure*}
    \centering
    \includegraphics[width=\linewidth]{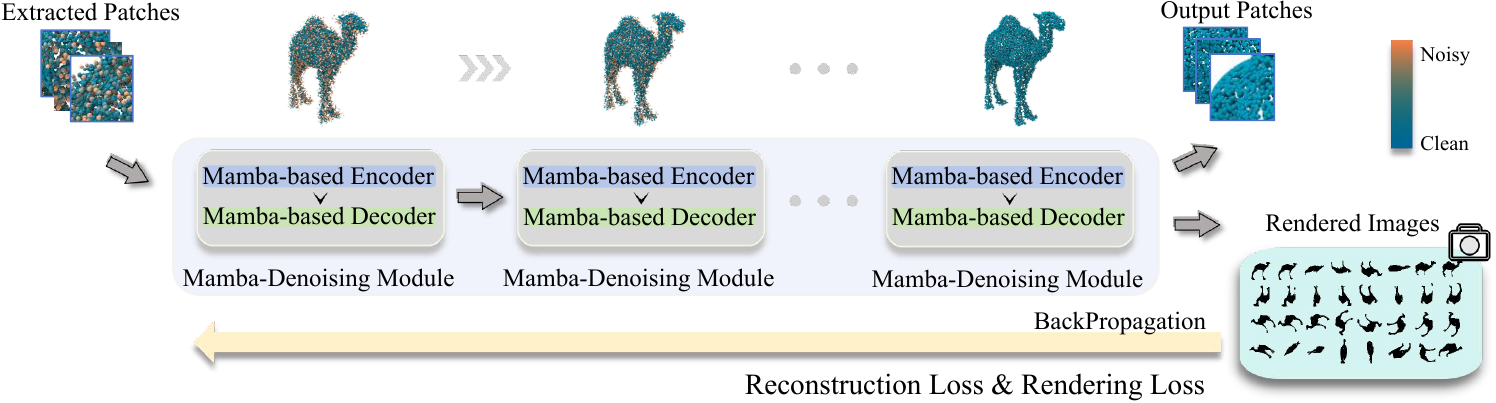}

    \caption{Overview of 3DMambaIPF. An Encoder-Decoder-based model built on Mamba named Mamba-Denoising Module is introduced for iterative filtering. The input noisy point cloud is partitioned into patches and fed into the iterative Mamba-Denoising Module. Upon completion of the iterations, clean point cloud patches are produced as output. To enhance the filtering of noisy points around the surface, a differentiable rendering method is introduced. The rendering loss and reconstruction loss are jointly backpropagated to update the parameters of 3DMambaIPF.}
    \label{fig:overview}

\end{figure*}

\textbf{Deep Learning-based Point Cloud Filtering Methods.} 
Deep Learning-based point cloud filtering methods are undergoing significant development to address the limitations of traditional approaches.
The well-known PCN (PointCleanNet)~\cite{rakotosaona2020pointcleannet} model is based on PCPNet~\cite{guerrero2018pcpnet} and focuses on removing outliers and reducing noise in unordered point clouds. 
However, due to PCN's denoising being performed point-wise rather than patch-wise, its efficiency is relatively low.
PD-Flow~\cite{mao2022pd} learns the distribution of noisy point sets and achieves denoising through noise disentanglement. 
However, the number of iterations needed for PD-Flow varies depending on the level of noise to achieve optimal results. 
GPDNet~\cite{pistilli2020learning} employs graph-convolutional layers to dynamically build neighborhood graphs and utilizes the similarity of high-dimensional feature representations to construct intricate feature hierarchies. 
ScoreDenoise~\cite{luo2021score} leverages the distribution model and score estimation from noisy point clouds.
Pointfilter~\cite{zhang2020pointfilter} employs an encoder-decoder architecture, where points and their neighbors are taken as input for point-wise learning. It generates a displacement vector to accomplish denoising while preserving sharp features.
IterativePFN~\cite{de2023iterativepfn} leverages iterative filtering modules and adaptive GT progressively yet swiftly converges noisy points onto clean surfaces, achieving superior performance.
Current point cloud filtering methods are primarily designed for sparse and small-scale datasets (up to 50K points) and struggle to effectively filter dense point clouds (approximately 500K points) with noise far from clean surfaces, particularly in large-scale point clouds.
Furthermore, existing methods \cite{zhang2020pointfilter} demonstrate limited effectiveness in filtering the edges of point clouds and capturing fine-grained characteristics.
Considering these challenges, our work aims to introduce a model tailored for highly dense and large-scale point clouds characterized by high-density noise and robust filtering capabilities at point cloud edges.

\textbf{SSM-based Methods.}\label{sec:ssm} 
SSMs provide a robust mathematical framework rooted in control theory and are increasingly applied in fields such as NLP and CV to model complex temporal dynamics within dynamic systems. 
This novel modeling approach addresses the challenge of handling long sequences, a difficulty encountered by both Transformers and RNNs. By mitigating the quadratic computational complexity of Transformers and addressing the forgetting issue of RNNs, SSMs prove highly efficient in capturing temporal dependencies within sequential data.


Significant advancements have been made in the Mamba framework~\cite{gu2023mamba}, which integrates selective SSMs into a simplified architecture. 
This integration allows Mamba to achieve rapid inference, linear scalability in sequence length, and state-of-the-art performance across multiple modalities, while also enhancing the selectivity of state representation. 
These advancements collectively demonstrate the ongoing evolution of SSMs and their increasing significance in modeling temporal dynamics across diverse domains. 
Methods based on Mamba have achieved state-of-the-art results in various fields, including large language models~\cite{he2024densemamba}, medical image analysis~\cite{liao2024lightm}, and 3D point cloud perception~\cite{liang2024pointmamba}.
Taking Consideration of the advantages of Mamba in modeling long sequences, we introduce Mamba in our work to address highly-dense and large-scale point cloud modeling problems.

\textbf{Differentiable Rendering Techniques.} 
Rendering techniques encompass algorithms employed to transform three-dimensional models into two-dimensional images.
The differentiable rendering methods are designed for introduction into deep learning, facilitating 3D reconstruction through error backpropagation.
Current differentiable rendering methods can be categorized into four types based on their geometric representations: point-based~\cite{yifan2019differentiable,muller2022unbiased,kerbl20233d,fei20243d}, triangle mesh-based~\cite{Laine2020diffrast,liu2019soft}, implicit function-based\cite{bangaru2022differentiable,vicini2022differentiable}, and volume-based\cite{mildenhall2021nerf, ren2023volrecon}.
In this paper, we propose an efficient differentiable rendering method based on point-based rendering~\cite{insafutdinov2018unsupervised} to render both the filtered and GT point clouds during training to generate the view loss, which is then propagated backward to adjust the network parameters.

\section{Method}
\subsection{Overview}

\begin{figure}[t]
    \centering
    
    \includegraphics[width=8cm]{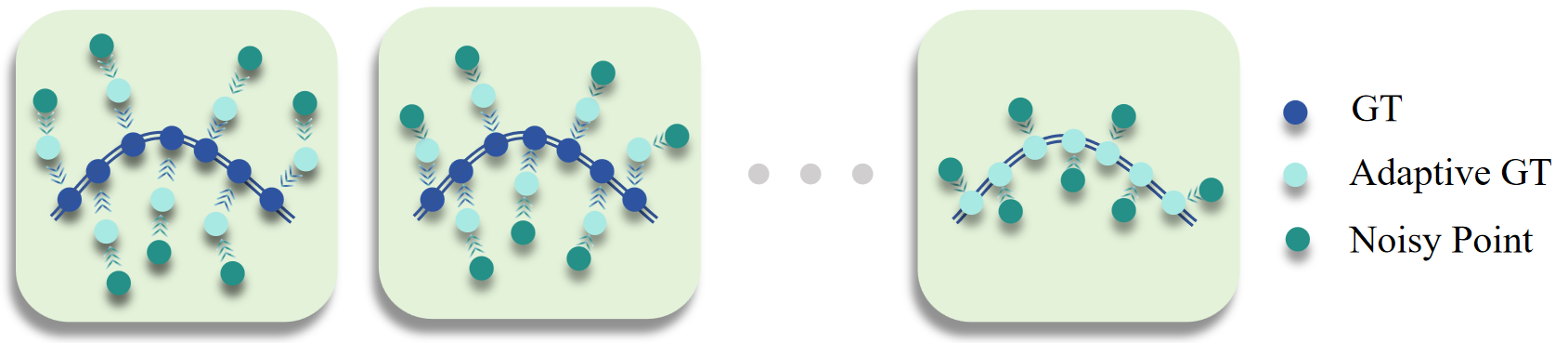}

    \caption{A Gaussian noise with a variable standard deviation is added to the GT to generate the adaptive GT. As the standard deviation gradually decreases, the adaptive GT approaches the GT with each iteration. Eventually, the GT no longer changes in the final iteration.}
    \label{fig:iter}

\end{figure}

\begin{figure}[t]
    \centering    \includegraphics[width=\linewidth]{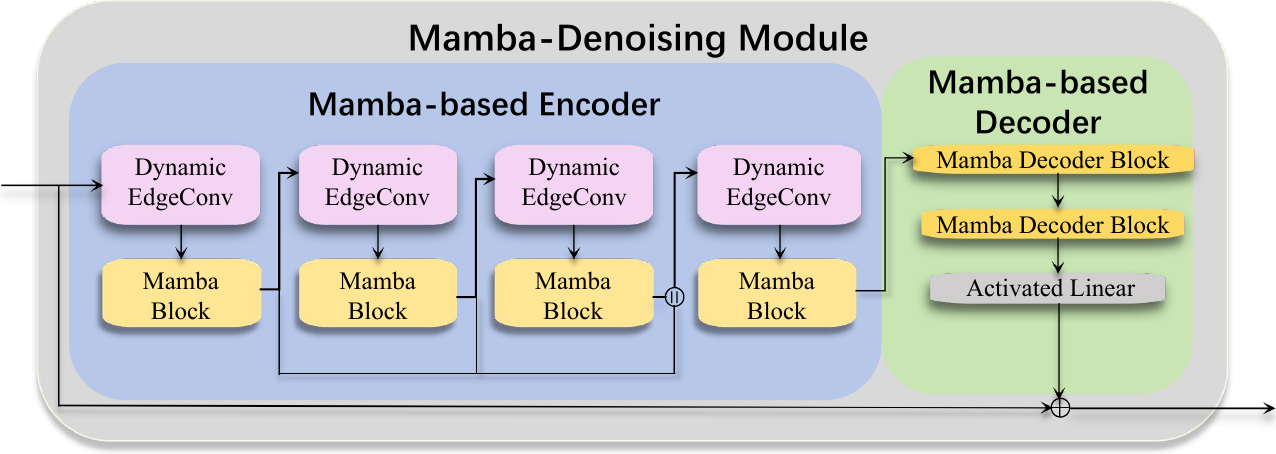}

    \caption{Mamba-Denoising Module for iterative point cloud filtering of 3DMambaIPF. The Mamba-based Encoder comprises Dynamic EdgeConv modules and Mamba Blocks, while the Mamba-based Decoder consists of Mamba Decoder Blocks and an Activated Linear module.}
    \label{fig:pcf}

\end{figure}

\begin{figure}
    \centering
    \includegraphics[width=\linewidth]{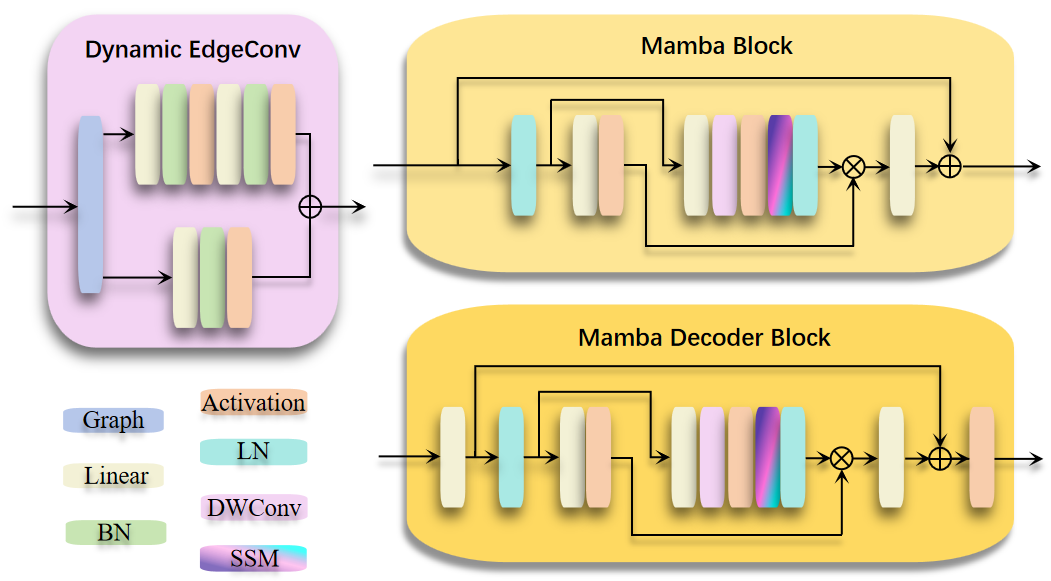}

    \caption{Detailed network architectures for Dynamic EdgeConv, Mamba Block, and Mamba Decoder Block.}
    \label{fig:blocks}

\end{figure}





The overview of 3DMambaIPF is depicted in Figure~\ref{fig:overview}, which utilizes a novel iterative Mamba-Denosing backbone and a differentiable point rendering technique.
Mamba-Denoising backbone consists of a cascade of Mamba-based encoders and decoders, with each pair referred to as a Mamba-Denoising Module.
After receiving an extracted patch of the point cloud as input, a single Mamba-Denoising Module constructs a directed graph from the points within the patch. 
The directed graph is encoded into token sequences to extract features. The denoised point token sequences are then decoded, eliminating noise, and serve as the input for the next Mamba-Denoising Module.
\begin{sloppypar}
Specifically, given a 3-dimensional clean point cloud, denoted as $X_c=\{x_{c1},x_{c2},...x_{cn}\}\subseteq\mathbb{R}^3$, each point in  $X_c$ can be represented by 3-dimensional coordinates $x_{ci}=(\alpha_i,\beta_i,\gamma_i)$. To prepare the point cloud as input for 3DMambaIPF, the K-nearest neighbors (KNN) algorithm~\cite{peterson2009k} is used to partition the point cloud into patches, and a directed graph is constructed within each patch. Considering a point cloud patch $X\subseteq\mathbb{K}^3$ which generated by KNN and a directed graph $\mathcal{G}=\{\mathcal{V},\mathcal{E}\}$, where $\mathcal{V}=\{x_1,x_2,...x_K\}$ and $\mathcal{E}$ represents the edges between 
$x_i\in\mathcal{V}$ and its $k$-nearest neighbors.
\end{sloppypar}

Simultaneously, the adaptive GT, inspired by the iterative point cloud filtering method~\cite{de2023iterativepfn}, is introduced to gradually pull noisy points closer to the clean surface with each iteration.
As shown in Figure~\ref{fig:iter}, GT is adaptive at each iteration, except for the last one. Specifically, to generate the adaptive GT $X_n\subseteq\mathbb{R}^3$, perturbed points are formed by adding Gaussian noise to $X_c$ with a standard deviation $\sigma$ between 0.5$\%$ and 2$\%$ of the radius of the bounding sphere. 
As the number of iterations increases, $\sigma$ decreases continuously, eventually reaching 0 at the final iteration. 
During each iteration, we aim to minimize the loss function to bring $x_i$ closer to its nearest neighbors in adaptive GT.

Subsequently, a differentiable 3D point rendering loss is employed to refine the positions of noisy points, especially around the surface of the point clouds, by comparing the rendered images from filtered point clouds with the adaptive GT.



\subsection{Mamba-Denoising Backbone}\label{sec:backbone}
The Mamba-Denoising Backbone is the core component of 3DMambaIPF for iterative point cloud filtering.
As illustrated in Figure~\ref{fig:pcf}, the Mamba-Denoising backbone adopts an Encoder-Decoder structure. The encoder consists of four Dynamic EdgeConv modules and four corresponding Mamba Blocks respectively, while the decoder comprises a Mamba Decoder Block and a linear activation layer. The detailed network structure is illustrated in Figure~\ref{fig:blocks}.

The Dynamic EdgeConv module facilitates the conversion of input patches into a graph structure, subsequently processed through an MLP network to extract pertinent features. Initially, vertices undergo feature extraction via an MLP, capturing intrinsic characteristics.
Subsequently, an additional MLP focuses on extracting positional features from vertices connected by edges to the target vertex, facilitating feature differentiation. 
Specifically, vertex features are updated as ${h_i}^{l+1} = f({h_i}^{l})+\sum_{i,j} g(\text{concat}({h_i}^{l},{h_j}^{l})-{h_i}^{l}),$ 
where $h^{l}$ denotes to features at layer $l$, $f(\cdot)$, $g(\cdot)$ denote one-layer or two-layers MLPs, $i$ is a vertex on the graph, $(i,j)$ forms an edge.
This process culminates in concatenating the target vertex's features with the differences of features computed from all connected vertices, effectively deriving nuanced positional features within the patch.
Following the feature extraction of the point cloud within the patch using MLPs, the resultant feature sequence is sent to the Mamba Encoder. This encoder selectively processes the input feature sequence and alters it into an output feature sequence.
In the Decoder module, the Mamba Decoder Block selectively processes the input feature sequence to reduce dimensionality, gradually reconstructing it into three dimensions. 
Linear activation is used to determine the output movement distance of a noisy point towards a clean surface after filtering. 
Utilizing a residual connection, the noisy points within the input patch are adjusted under the guidance of the movement distance, resulting in a denoised patch.

\subsection{Differentiable Rendering and Backpropagation}\label{sec:back}

For enhanced filtering of the noisy points around the surface, we propose a differentiable rendering method and integrate it in 3DMambaIPF to explicitly quantify the level of noise and enable backpropagation.
To render the point clouds, the 3D coordinates of the raw point clouds are first converted into the standard coordinate frame via a projective transformation aligned with the camera pose.
Subsequently, each discretized point is expressed as scaled Gaussian densities, yielding the occupancy map, and making the backpropagation available. Through the introduction of a differentiable ray tracing operator, these occupancies are transformed into ray termination probabilities. Ultimately, the rendered image is generated by projecting the volume onto the plane.

Denoised point cloud and adaptive GT are utilized to produce $K$ pairs of images with fixed camera poses using point rendering technique introduced above. 
Our differentiable rendering method employs three projection planes, allowing for the rendering of images with greater detail and coverage of the overall shapes, rather than projecting them onto a single plane.

The rendering loss is defined in Equation~\ref{renderloss}. It aims to characterize the local distribution of points and penalize noise distribution, encouraging the denoised point cloud to achieve an improved geometric appearance, particularly around the surfaces of the point cloud, throughout the backpropagation process:

\begin{equation}\label{renderloss}
\begin{aligned}
{\mathcal{L}}_{Render}(t)=\frac{1}{K} \sum_{i=1}^K  \lvert V_i(\mathcal{P}_t) - V_i(\hat{\mathcal{P}}) \rvert,
\end{aligned}
\end{equation}
where $\hat{\mathcal{P}}$ denotes the predicted point cloud, $\mathcal{P}_t$ denotes the adaptive GT at the $t$-th iteration, $K$ is the quantity of rendered views, and $V_i$ represents a rendered view in a specific camera position.

\subsection{Loss Function}\label{sec:loss}
Our loss function comprises two components: reconstruction loss and rendering loss. 
The reconstruction loss aims to minimize the distance between each point in the noisy point cloud and its closest counterpart in the adaptive GT, thereby structurally denoising the point cloud. 
Specifically, it seeks to bring every point in the noisy point cloud closer to its corresponding (the nearest) point in the adaptive GT.

Due to the possibility of overlapping regions within each patch during patch partitioning, we introduce a patch stitching method~\cite{zhou2022fast}. This method weights input points based on the proximity of each point $p_i$ to a reference point $p_r$ with a Gaussian distribution.
Specifically, the weight $w_i$ for each $p_i$ is calculated as:

\begin{equation}\label{wi}
\begin{aligned}
w_i = \frac{\exp(-{\lVert p_i-p_r \rVert}^2 /2{r_s}^2)}{\sum_i\exp(-{\lVert p_i-p_r \rVert}^2 /2{r_s}^2)},
\end{aligned}
\end{equation}
where the support radius $r_s$ is set to $r/3$, and $r$ denotes to the patch radius. Then, the reconstruction loss at the $t$-th iteration can be defined as:

\begin{equation}\label{reconloss}
\begin{aligned}
{\mathcal{L}}_{Recon}(t)= \sum_{\hat{p_i}\in \hat{\mathcal{P}}} w_i \min_{{p_{ti}} \in \mathcal{P}_t}{{\lVert p_{ti} - \hat{p_i} \rVert}^2_2},
\end{aligned}
\end{equation}
where $\hat{\mathcal{P}}$ denotes the predicted point cloud, $\mathcal{P}_t$ denotes the adaptive GT at the $t$-th iteration, and $p_i$ represents a point in point cloud $\mathcal{P}$.

Loss function of one iteration is a weighted combination of the reconstruction loss and the rendering loss, and the ultimate loss is obtained by aggregating loss contributions across iterations, which are summed up as:

\begin{equation}\label{totalloss}
\begin{aligned}
{\mathcal{L}} = \sum_{t=1}^T({\mathcal{L}}_{Recon}(t)+\alpha{\mathcal{L}}_{Render}(t)),
\end{aligned}
\end{equation}
where $T$ is number of iteration times, and weight $\alpha$ is 0.01 in experiments. The detailed pseudocode for training 3DMambaIPF is provided in Appendix.

\section{Experiments}
\begin{figure*}[t]
    \centering
    \includegraphics[width=\linewidth]{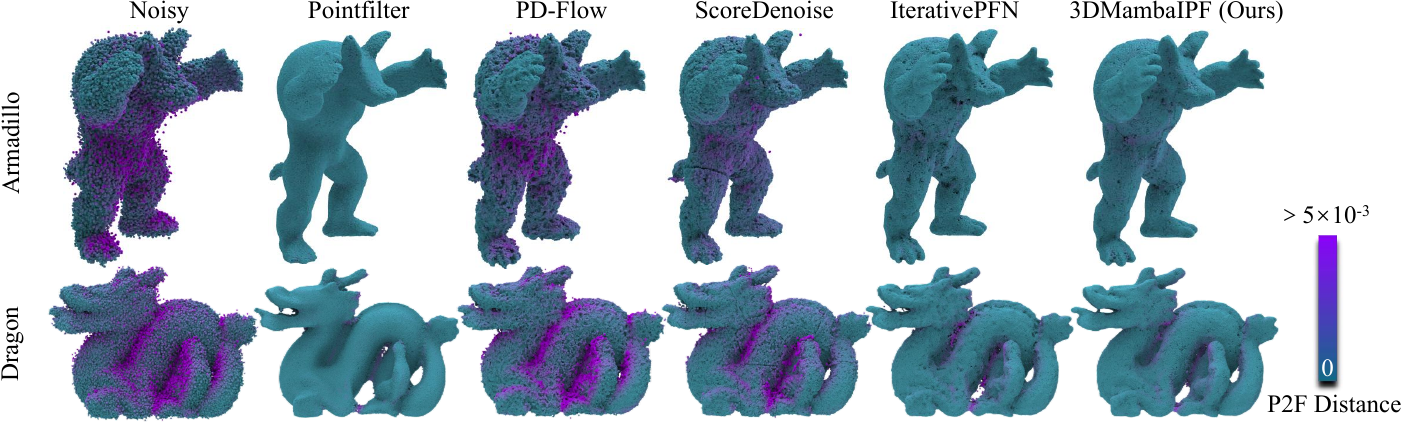}

    \caption{Visualization Comparisons on Stanford 3D Scanning Repository with 5$\textbf{\%}$ Gaussian deviation. Coloration of each point is determined by its point-wise P2F distance, with points exhibiting low P2F values (clean) depicted in green, while those with high P2F values (noisy) are portrayed in purple.}
    \label{fig:viewdragon}

\end{figure*}

\begin{figure*}[t]
    \centering
    \includegraphics[width=\linewidth]{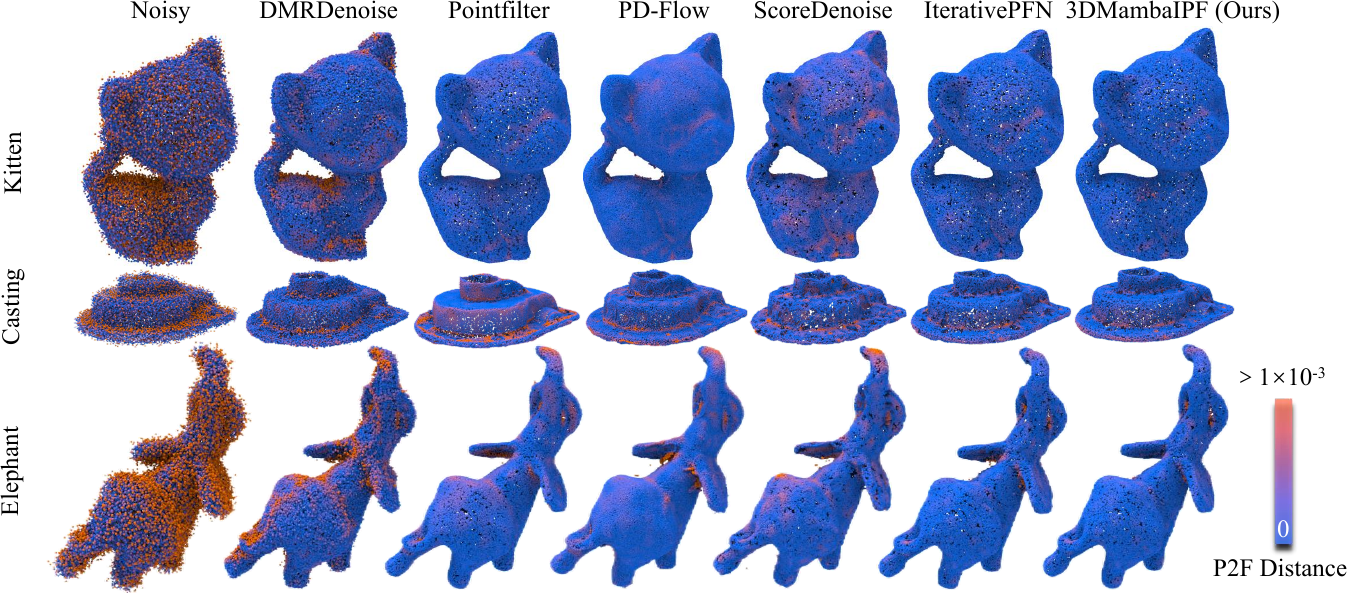}
    \caption{Visualization Comparisons on PU-Net Dataset of 50K points with 2.5$\textbf{\%}$ Gaussian deviation. The coloration of each point is determined by its point-wise P2F distance, with points exhibiting low P2F values (clean) depicted in blue, while those with high P2F values (noisy) are portrayed in red.}
    \label{fig:viewpunet}
\end{figure*}


\begin{table}[t]

\centering
\resizebox{\linewidth}{!}{
\begin{tabular}{c|cccc}
\toprule [1.5pt]
\multirow{3}{*}{Method}

&\multicolumn{2}{c}{Armadillo}
&\multicolumn{2}{c}{Dragon}

  \\\cline{2-5}
  &\multicolumn{2}{c}{173K points}&\multicolumn{2}{c}{438K points}\\ 
\cline{2-5}
&CD&P2M&CD&P2M\\ \toprule [1pt]
Noisy &50.19 &47.96&38.92&37.51\\
Pointfilter &\textbf{6.41}&\textbf{4.82}&\textbf{12.10}&\textbf{12.70}\\
PD-Flow &52.30&50.08&51.48&50.00\\
ScoreDenoise &29.42&27.41&36.25&34.88\\

IterativePFN &\underline{17.04}&\underline{14.93}&20.43&18.98\\\toprule [1pt]
3DMambaIPF (Ours)& 17.14&15.06&\underline{19.79}&\underline{18.36}\\\toprule [1.5pt]
\end{tabular}
}

\caption{Results on Stanford 3D Scanning Repository. CD and P2M distances are multiplied by $10^{5}$.}
\label{exp_stan}

\end{table}

\subsection{Overview of Comparative Baselines}
The filtering capabilities of comparative baselines are conducted with DMRDenoise~\cite{luo2020differentiable}, Pointfilter~\cite{zhang2020pointfilter}, PD-Flow~\cite{mao2022pd}, ScoreDenoise~\cite{luo2021score} and IterativePFN~\cite{de2023iterativepfn}.
IterativePFN and 3DMambaIPF work on highly-dense and large-scale datasets.
Pointfilter tends to obtain over-smoothed surfaces that deviate from the GT and fail to keep the original structure after filtering and struggles to work on highly-dense point clouds.
On the other hand, the point cloud filtering task is not addressed as a \textbf{multi-modal} issue in previous works.
All the baselines focus solely on either the movement of individual noise points or the generation of clean points, which leads to cluttered visual results, particularly around the surface of the point clouds.

\subsection{Datasets and Experimental Settings}





Details about the PU-Net~\cite{yu2018pu} dataset can be found in Appendix. Due to the limited number of points in PU-Net dataset, which contains a maximum of 50K points, point clouds from the Stanford 3D Scanning Repository~\cite{curless1996volumetric,krishnamurthy1996fitting} are utilized to evaluate the filtering performance of large-scale synthetic objects.
The preprocessing of datasets is detailed in the Appendix.

The experiments are conducted on NVIDIA A100 GPUs, and 3DMambaIPF is implemented with PyTorch 1.13.1 and CUDA 11.7.
The Mamba modules are applied with 6 layers, and the total parameters of Mamba-Denoising Modules is 0.81M. 
Following previous works~\cite{luo2021score,de2023iterativepfn}, the quantitative indicators CD and P2M are adopted for evaluation and comparison.
More details can be found in Appendix.

\begin{table*}[t]

\centering
\resizebox{\linewidth}{!}{
\begin{tabular}{c|cccccc|cccccc}

\toprule [1.5pt]
\multirow{3}{*}{Method}
&\multicolumn{6}{c|}{10K points}
&\multicolumn{6}{c}{50K points}\\ \cline{2-13}
&\multicolumn{2}{c}{1$\%$ noise}
&\multicolumn{2}{c}{2$\%$ noise}
&\multicolumn{2}{c|}{2.5$\%$ noise}
&\multicolumn{2}{c}{1$\%$ noise}
&\multicolumn{2}{c}{2$\%$ noise}
&\multicolumn{2}{c}{2.5$\%$ noise}  \\ \cline{2-13}
&CD&P2M&CD&P2M&CD&P2M&CD&P2M&CD&P2M&CD&P2M \\ \toprule [1pt]
Noisy &36.90&16.03&79.39&47.72&105.02&70.03&18.69&12.82&50.48&41.36&72.49&62.03\\
PCN   &36.86&15.99&79.26&47.59&104.86&69.87&11.03& 6.46&19.78&13.70&32.03&24.86\\
GPDNet&23.10&7.14&42.84&18.55&58.37&30.66&10.49&6.35 &32.88& 25.03& 50.85& 41.34\\
DMRDenoise& 47.12& 21.96 &50.85 &25.23 &52.77& 26.69 &12.05& 7.62 &14.43& 9.70 &16.96& 11.90\\
Pointfilter &24.61 &7.30& 35.34& 11.55& 40.99 &15.05& 7.58& 4.32& 9.07 &5.07 &10.99& 6.29\\
PD-Flow &21.26 &6.74& 32.46 &13.24 &36.27& 17.02& 6.51 &4.16& 12.70 &9.21& 18.74 &14.26\\
ScoreDenoise &25.22& 7.54 &36.83& 13.80 &42.32 &19.04 &7.16 &4.00 &12.89& 8.33 &14.45 &9.58\\
IterativePFN &20.55& 5.01& 30.43& 8.43& 33.53& 10.46& 6.05 &3.02 &8.03&4.36 &10.15& 5.88\\\toprule [1pt]
3DMambaIPF (Ours)& \textbf{19.89}& \textbf{4.77} &\textbf{29.95}&\textbf{8.03}   &\textbf{32.62}&\textbf{9.92}& \textbf{5.89}&\textbf{2.91} &\textbf{7.55} &\textbf{4.05}  &\textbf{9.28} &\textbf{5.31}\\\toprule [1.5pt]
\end{tabular}
}

\caption{Results on PU-Net Dataset. CD and P2M distances are multiplied by $10^{5}$.}
\label{exp1}
\end{table*}

\begin{table}[t]

\centering
\resizebox{\linewidth}{!}{
\begin{tabular}{c|c|cccccc}
\toprule [1.5pt]
\multirow{3}{*}{\makecell[c]{Rendering\\ Views}}
&\multirow{3}{*}{\makecell[c]{Mamba\\ Layers}}
&\multicolumn{6}{c}{50K points}\\ \cline{3-8}
&
&\multicolumn{2}{c}{1$\%$ noise}
&\multicolumn{2}{c}{2$\%$ noise}
&\multicolumn{2}{c}{2.5$\%$ noise}  \\ \cline{3-8}
&&CD&P2M&CD&P2M&CD&P2M \\ \toprule [1pt]

/ &/& 6.05 &3.02 &8.03&4.36 &10.15& 5.88\\\toprule [1pt]
/ &6&5.94&2.92&7.70&4.14&9.41&5.37\\

8& 6    &6.04  &2.99 & 8.06 &4.36  & 10.39 &6.05   \\

 16 &6  &6.21  &3.14 &7.87  & 4.29&  9.63& 5.60  \\

24&6  &5.96  &2.97 &7.81  &4.26  &9.56  & 5.53  \\

 32&6 & \textbf{5.89}&\textbf{2.91} &\textbf{7.55} &\textbf{4.05}  &\textbf{9.28} &\textbf{5.31}\\\midrule [1pt]

 32 & 1 &8.15  &4.71  &10.82  &6.53   & 13.23&8.20  \\

32& 3 & 6.12& 3.09 &  8.05&  4.43& 9.58&5.54  \\

32& 6 & \textbf{5.89}&\textbf{2.91} &\textbf{7.55} &\textbf{4.05}  &\textbf{9.28} &\textbf{5.31}\\

32 & 9 &6.15  &3.08 & 8.01 & 4.38 &  9.89&5.74   \\

32 & 12   &6.12  &3.07 & 8.09 &4.42  & 9.85 & 5.67  \\\bottomrule [1.5pt]

\end{tabular}
}

\caption{Ablation results of Mamba layers and rendered views on PU-Net 50K dataset. CD and P2M distances are multiplied by $10^{5}$.}
\label{total_ablation}
\end{table}
\subsection{Comparisons on Highly-dense and Large-scale Datasets}
The filtering performance of highly dense and large-scale point clouds is evaluated on synthetic models from the Stanford 3D Scanning Repository.
The quantitative and visual results are shown in Table~\ref{exp_stan} and Figure~\ref{fig:viewdragon}. 
Due to the presence of dense and distant noise relative to the surface, some baselines struggle to effectively filter the point clouds.
While Pointfilter achieves the best quantitative results, it lacks significant geometric details and accurate surface structures.
It is found that the incorporation of the generative model into PD-Flow results in varied denoising outcomes across different experiments on the same point cloud.
Despite potential variations in outcomes, PD-Flow generally falls short in effectively denoising highly-dense point clouds, particularly in severely noisy areas where it struggles to achieve significant noise reduction.
ScoreDenoise, employing a patch partitioning method, exhibits gaps in results when processing highly-dense and large-scale point clouds. 
Although IterativePFN tackles this issue with patch-stitching and achieves competitive quantitative results, its performance is limited when dealing with point cloud surfaces and densely cluttered areas.
As a single-modal filtering method, it struggles with the visual quality of point cloud surfaces, leading to gaps on the object surface that do not align with GT.
On the contrary, 3DMambaIPF achieves the best performance when dealing with highly dense and large-scale point clouds.
Additionally, the utilization of rendering loss results in significantly improved filtering effects and better preservation of geometric details, particularly around the surfaces of the point clouds.

\subsection{Comparisons on PU-Net Dataset}\label{sec:smallexp}
Table~\ref{exp1} presents the results of 
3DMambaIPF compared to other baseline methods on PU-Net dataset. 
Figure~\ref{fig:viewpunet} illustrates the visualization comparison on PU-Net dataset.
Each visualized object contains 50K points with a Gaussian deviation of 2.5$\%$ of the bounding sphere radius.
3DMambaIPF outperforms all baseline methods across different resolutions and various levels of noise.
Moreover, visual results have shown that 3DMambaIPF exhibits superior filtering performance, particularly with high accuracy in boundary denoising.
Specifically, while Pointfilter demonstrates competitive results in terms of CD and P2M metrics, it exhibits geometrically unrealistic deviations in boundary denoising for complex objects (e.g. casting).
Compared to other baselines, IterativePFN excels both in terms of metrics and visual presentation, with its only drawback being slightly inferior denoising effects on point cloud boundaries.
Conversely, 3DMambaIPF ensures the authenticity of geometric structures and exhibits superior denoising results for boundary sections, while also outperforming in metrics.


\section{Ablation Study}

\subsection{Ablation on Modules}
Ablation experiments are firstly conducted on the influencing factors of 3DMambaIPF, focusing on two aspects: i) the quantity of rendered images in point rendering, ii) the number of Mamba layers. 
Table~\ref{total_ablation} validates the enhancement of the filtering capability of 3DMambaIPF across various levels of noise with the introduction of view loss, and proceeds to examine how varying numbers of rendered views affect filtering effectiveness.
However, employing an excessive number of views will decrease the training speed, and detailed runtime metrics will be validated in appendix.
Subsequently, ablation experiments are carried out on the number of Mamba layers. 
Table~\ref{total_ablation} illustrates that the denoising capability is at its peak with 6 Mamba layers.

\subsection{Runtime of Differentiable Rendering}
An additional ablation for the runtime of rendered views is conducted as shown in Table~\ref{runtime}.
Runtime$_{DR}$ represents the total time required to render both the GT and the predicted point clouds. 
Runtime$_{Step}$ denotes the total time for one training step.
Due to our devised network-free differentiable rendering, the runtime of differentiable rendering only utilizes 11.8\%-16.5\% of the total time for one training step.
Except for this efficiency, differentiable rendering demonstrates notable improvements of 12.0\% and 13.9\% in CD and P2M metrics, respectively, as the number of views increases. This trade-off between runtime and performance is cost-effective.

\begin{table}[t]

\centering
\resizebox{\linewidth}{!}{
  \begin{tabular}{c|ccccc}
   \toprule[1.5pt]
  \multirow{2}{*}{Views}&\multicolumn{4}{c}{Metrics}\\
&CD&P2M&Runtime$_{DR}$ (s)&Runtime$_{Step}$ (s)\\
    \midrule[1pt]

    8&10.39&6.05&0.08&0.74\\
    16&9.63&5.60&0.11&0.82\\
    24&9.56&5.53&0.14&0.97\\
    32&9.28&5.31&0.17&1.03\\\bottomrule[1.5pt]
  \end{tabular}}
  \caption{Runtime analysis on differentiable rendering. CD and P2M metrics are multiplied by $10^5$, and tested on PU-Net Dataset of 50K points with 2.5\% Gaussian deviation.}
  \label{runtime}
\end{table}

\subsection{Comparisons 3D Mamba with Other Backbones}
To validate the effectiveness of our designed 3D Mamba, ablation studies (Table~\ref{modules}) comparing the Mamba modules with Transformers and MLPs are conducted. 
For fairness, the number of modules, layers and embedding sizes are consistent with 3DMambaIPF.
For intuitive results, differentiable rendering loss is not implemented in this ablation.
3DMambaIPF fulfills the best denoising performance while ensuring fast training speed as well as the smallest model size.

\begin{table}[t]

\centering
\resizebox{\linewidth}{!}{
  \begin{tabular}{l|ccccc}
   \toprule[1.5pt]
  \multirow{2}{*}{Methods}&\multicolumn{5}{c}{Metrics}\\
&CD&P2M&GFLOPs&FPS&MParams\\
    \midrule[1pt]

    with Transformers &\underline{14.02}&\underline{8.56}&178.76&2.31&158.00\\
    with MLPs& 30.17&21.71&\underline{20.78}&\textbf{4.39}&\underline{4.56} \\
    with Mambas (Ours) &\textbf{9.42}&\textbf{5.37}&\textbf{16.92}&\underline{3.62}&\textbf{0.81}\\\bottomrule[1.5pt]
  \end{tabular}   }

  \caption{Comparsions on Mamba with other backbones. 
  CD and P2M metrics are multiplied by $10^5$, and tested on PU-Net Dataset of 50K points with 2.5\% Gaussian deviation.
  }
  \label{modules}
\end{table}

\section{Conclusion}
In this study, we introduce 3DMambaIPF, a novel iterative Mamba-based point cloud filtering model that utilizes a differentiable rendering technique for the first time in this area. 3DMambaIPF not only achieves state-of-the-art results on a commonly-used dataset but also demonstrates superior performance on a highly dense and large-scale dataset, indicating its capability to address denoising challenges in extensive scenes, as validated through ablation experiments. Although 3DMambaIPF may encounter reverse-amplification issues on denoising a smooth and clean surface, its performance remains optimal.
In future work, we will explore different datasets to address the challenges of filtering real-world point clouds from large scenes.

\section{Acknowledgments} This project was partially supported by the National Natural Science Foundation of China (U2033209); Hong Kong Innovation and Technology Fund (PiH/213/24GS); the Ministry of Education, Singapore, under its Academic Research Fund Grants (MOE-T2EP20220-0005 \& RT19/22) and an OPPO gift fund.
\bibliography{ref}

\newpage
\newpage
\clearpage



\maketitle

\section{Preliminaries}

\textbf{Structured State Space Models} are two-stage sequence-to-sequence models that map the input $x(t)\in\mathbb{R}^L$ to output $y(t)\in\mathbb{R}^L$. 
Specifically, the SSM process can be described as follows:
\begin{equation}
\begin{aligned}
    h'(t)=\bf{A}&h(t)+{\bf{B}}x(t),  \\
       y(t)&={\bf{C}}h(t),     
\end{aligned}
\end{equation}
where $h(t)\in\mathbb{R}^N$ represents the latent state, and $h'(t)\in\mathbb{R}^N$ is the derivative of $h(t)$. ${\bf{A}}\in\mathbb{R}^{N\times{N}}$, ${\bf{B}}\in\mathbb{R}^{N\times{1}}$, ${\bf{C}}\in\mathbb{R}^{1\times{N}}$ are the parameters.
The first stage of S4 discretizes the parameter ${\bf{A}}$, ${\bf{B}}$ into ${\bf{\bar{A}}}$, ${\bf{\bar{B}}}$ with a parameter $\Delta$:
\begin{equation}
\begin{aligned}
    {\bf{\bar{A}}}=\exp&(\Delta {\bf{A}}),\\  
    {\bf{\bar{B}}}=(\Delta {\bf{A}})^{-1}(\exp&(\Delta {\bf{A}})-I)\cdot{\Delta {\bf{B}}}.
\end{aligned}
\end{equation}
After discretizing, the second stage is to calculate recursively:
\begin{equation}
\begin{aligned}
    h_t={\bf{\bar{A}}}&h_{t-1} + {\bf{\bar{B}}}x_t,\\  
    y_t&={\bf{C}}h_t,
\end{aligned}
\end{equation} with the convolution formula:
\begin{equation}
\begin{aligned}
    {\bf{\bar{K}}}=({\bf{C}}\bf{\bar{B}},{\bf{C}}{\bf{\bar{A}}}{\bf{\bar{B}}}&,...,{\bf{C}}{\bf{\bar{A}}}^k{\bf{\bar{B}}},...),\\
    y=x&*{\bf{\bar{K}}}.
\end{aligned}
\end{equation}

\textbf{Mamba} employs recurrent scans coupled with a selection mechanism to regulate the passage of specific segments within the sequence into the latent states. 
Specifically, by making the SSM parameters functions of the input, Mamba addresses the weakness in dealing with discrete modalities, allowing the model to selectively propagate or forget information based on the current token. 
Simultaneously, Mamba has devised a hardware-aware parallel algorithm to operate in recursive mode, addressing the challenge of inefficient convolution operations. 
The overall process of a single Mamba block can be viewed as a transition from $x_{t-1}\in\mathbb{R}^L$ to $x_t\in\mathbb{R}^L$:
\begin{equation}
\begin{aligned}
{x_t}'=\text{DWConv}(\text{MLP}&(\text{LN}(x_{t-1}))),\\
s_t =\text{MLP}(\text{LN}(\text{SSM}(\sigma({x_t}&')))\times\sigma(\text{LN}(x_{t-1}))),\\
x_t=s_t+x&_{t-1},
\end{aligned}
\end{equation}where $\text{DWConv}$ represents depth-wise convolution, $\text{LN}$ represents layer normalization, $\sigma$ denotes activation (SiLU here).

\section{Training Details}
\subsection{Pseudocode}
The pseudocode for training 3DMambaIPF is shown in Algorithm 1.
\begin{algorithm}[h]
    \SetAlgoLined 
    \label{al1}
	\caption{The pseudocode for training 3DMambaIPF. All of the parameters are introduced in Implementation Details.}
	\KwIn{noisy point cloud $PCD_{noisy}$, ground truth point cloud $PCD_{GT}$, number of iteration times $T$, number of Mamba-Denoising Modules $M$, weight of rendering loss $\alpha$.}
 
	\KwOut{filtered point cloud $PCD_{predict}$, loss $L$.}
        
        $L$ = 0 
        
        $P$ = \text{Extract} ($PCD_{noisy}$) \# Patch Extracting\;
        
	\For{i from 1 to T}{
             $PCD_{ad\_GT}$ = \text{Add\_noise} ($PCD_{GT}$, $i$)    \#Adaptive GT\;
             
             \For{m \text{from 1 to} M}{
             
             $P$ = \text{Mamba\_Denoising} ($P$)  \#Filtering
		}
            $PCD_{predict}$ = \text{Stitch} ($P$) \# Patch Stitching\;
            
            $L_{recon}$ = \text{Recon\_loss} ($PCD_{predict}$, $PCD_{ad\_GT}$) 
            
            $L_{render}$ = \text{Render\_loss} ($PCD_{predict}$, $PCD_{ad\_GT}$)

            $L$ += $L_{recon} + \alpha \times L_{render}$ 
	}

	\textbf{return} $L$
\end{algorithm}

\subsection{Rendering Loss and Backpropagation}
\begin{figure*}[t]
    \centering
    \includegraphics[width=\linewidth]{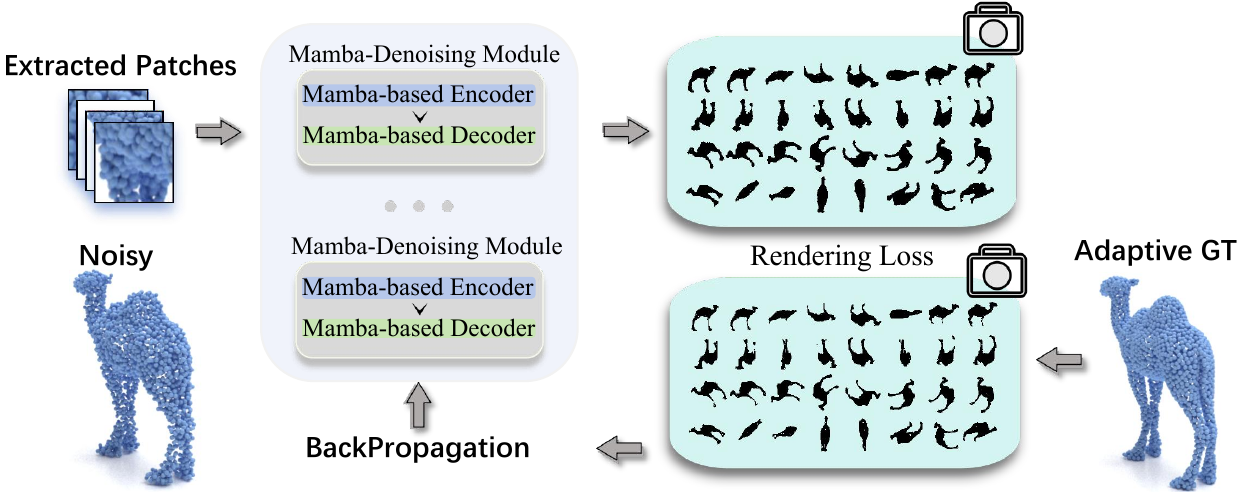}

    \caption{Diagram for rendering loss and backpropagation.}
    \label{fig:diff}

\end{figure*}

Figure~\ref{fig:diff} shows the diagram for rendering loss and backpropagation.

\begin{figure*}[h]
    \centering
    \includegraphics[width=\linewidth]{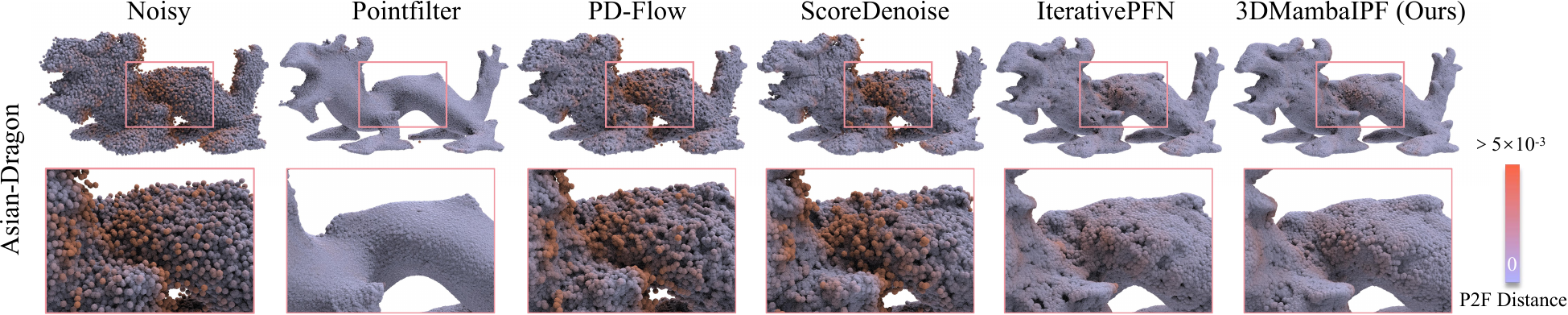}

    \caption{Visualization Comparisons on Asian-Dragon from the Stanford 3D Scanning Repository with 5$\textbf{\%}$ Gaussian deviation. Coloration of each point is determined by its point-wise P2F distance, with points exhibiting low P2F values (clean) depicted in purple, while those with high P2F values (noisy) are portrayed in red.}
    \label{fig:viewasiandragon}

\end{figure*}

\begin{figure*}[h]
    \centering
    \includegraphics[width=\linewidth]{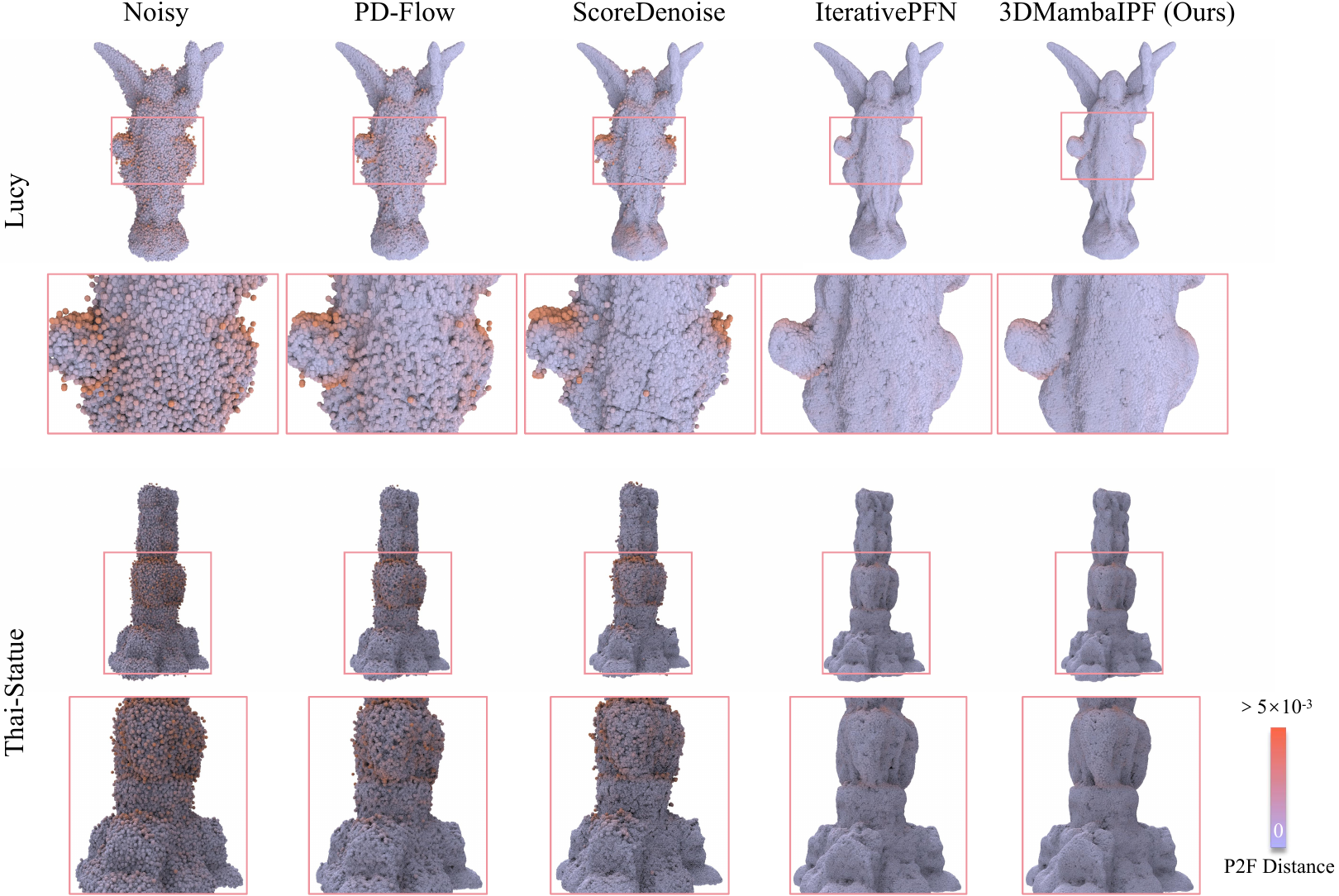}

    \caption{Visualization Comparisons on Lucy and Thai-Statue from the Stanford 3D Scanning Repository with 5$\textbf{\%}$ Gaussian deviation. Coloration of each point is determined by its point-wise P2F distance, with points exhibiting low P2F values (clean) depicted in purple, while those with high P2F values (noisy) are portrayed in red.}
    \label{fig:viewthailucy}

\end{figure*}

\begin{figure*}[h]
    \centering
    \includegraphics[width=\linewidth]{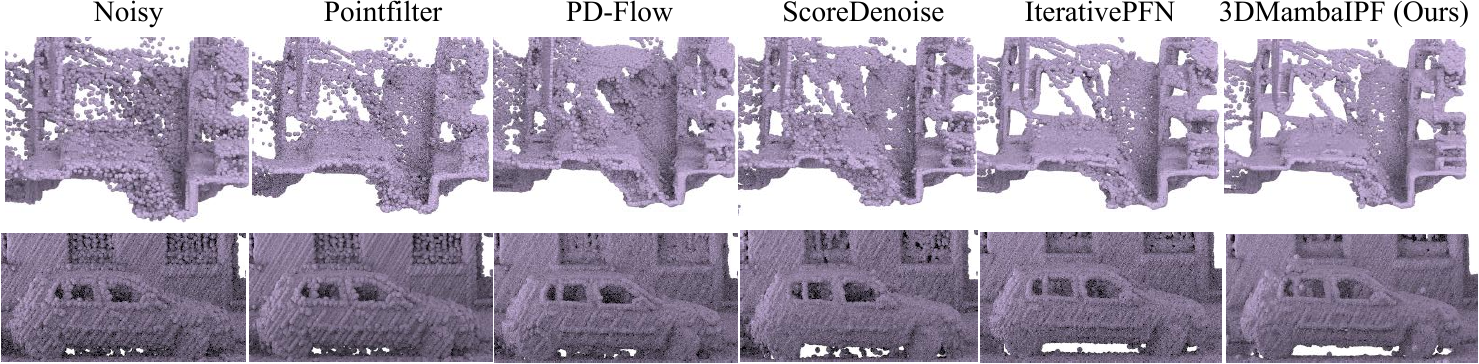}

    \caption{Visualization Comparisons on Paris-rue-Madame dataset.}
    \label{fig:viewrue}

\end{figure*}

\section{Detailed Experiments Settings}
\begin{table*}[ht]

\centering
\tabcolsep=0.07cm
\resizebox{\linewidth}{!}{
  \begin{tabular}{c|cccccc}
   \toprule[1.5pt]
Methods&DMRDenoise&Pointfilter&PD-Flow&ScoreDenoise&IterativePFN&3DMambaIPF\\

    \midrule[1pt]
Highly-dense \& large-scale   & $\usym{2718}$& $\usym{2718}$ & $\usym{2718}$& $\usym{2718}$& $\usym{2714}$&$\usym{2714}$\\
Multi-modal    & $\usym{2718}$& $\usym{2718}$ & $\usym{2718}$& $\usym{2718}$& $\usym{2718}$& $\usym{2714}$\\
Original pattern & $\usym{2714}$& $\usym{2718}$ & $\usym{2714}$ & $\usym{2714}$& $\usym{2714}$& $\usym{2714}$\\\bottomrule[1.5pt]

\end{tabular}}

  \caption{Technical comparison with off-the-shelf methods.}
\label{compare_tech}

\end{table*}
\subsection{PU-Net Dataset}
Following previous works~\cite{mao2022pd,luo2021score,de2023iterativepfn}, 3DMambaIPF is trained with PU-Net~\cite{yu2018pu} dataset.
The training set consists of 40 sets of meshes, from which point clouds are generated at resolutions of 10K, 30K, and 50K points, resulting in a total of 120 point clouds for training.
To create noisy inputs, Gaussian noise is added to each point cloud with standard deviations of 0.5$\%$, 1.0$\%$, 1.5$\%$, and 2$\%$ of the diagonal length of the object bounding box respectively. Similar to IterativePFN, we conduct testing with 20 sets of meshes, generating point clouds at resolutions of 10K and 50K, resulting in a total of 40 point clouds.

\subsection{Stanford 3D Scanning Repository}
For experimentation, five large-scale synthetic objects have been selected from the Stanford 3D Scanning Repository~\cite{krishnamurthy1996fitting,curless1996volumetric}, Dragon, Armadillo, Asian-Dragon, Lucy, and Thai-Statue. 
Objects with fewer than 500K points retain all the points, while those with more than 500K points have 500K points randomly sampled. 
Gaussian noise with a standard deviation of 5$\%$ of the diagonal length of the object bounding box is added to simulate the noisy inputs.
Similar to PU-Net, the GT mesh retains all the points and faces of the original object, while the GT point clouds are sampled and contain no more than 500K points.

\subsection{Implementation Details} 
The 3DMambaIPF network is trained on NVIDIA A100 GPUs with PyTorch 1.13.1 and CUDA 11.7. 
The network underwent 100 epochs of training on PU-Net dataset, employing the Adam optimizer with a learning rate of 1 × $10^{-4}$. The number of Mamba-Denoising Modules is 4, and the number of iteration times is 4. We apply Mamba modules with 6 layers, and the total parameters of Mamba-Denoising Modules is 0.81M. Patch size is set to 2000 in order to process larger point clouds. As mentioned in the main text, the weight of rendering loss is 0.01.

For consistency across evaluations, all experimental methods (including 3DMambaIPF) are evaluated on the same devices in this paper.
For a small-scale dataset (PU-Net), all experimental methods (including 3DMambaIPF) are evaluated on a single NVIDIA GeForce RTX 3090 GPU.
For a large-scale dataset (point clouds from the Stanford 3D Scanning Repository), all experimental methods (including 3DMambaIPF) are tested on a single NVIDIA A100 GPU.

\subsection{Evaluation Metrics}
Referring to many prior notable works, we utilize Chamfer Distance (CD) and Point-to-Mesh (P2M) Distance as the evaluation metrics for our study. Notably, CD is delineated in Equation~\ref{eq4.1}, encompassing the combined sum of two components: the average distances between each point in the predicted point cloud and its nearest point in the GT point cloud, and the average distances between each point in the GT point cloud and its nearest point in the predicted point cloud.

\begin{equation}\label{eq4.1}
\begin{aligned}
{\mathcal{L}}_{CD}(\hat{\mathcal{P}},\mathcal{P})=\frac{1}{\vert\hat{\mathcal{P}}\vert}\sum_{{\hat{p}\in\hat{\mathcal{P}}}} \min_{{p} \in \mathcal{P}} \Vert \hat{p}-{p}\Vert + \frac{1}{\vert\mathcal{P}\vert}\sum_{{p} \in \mathcal{P}} \min_{{\hat{p}\in\hat{\mathcal{P}}}} \Vert {p}-\hat{p}\Vert,
\end{aligned}
\end{equation}
where $\hat{\mathcal{P}}$ represents the predicted point cloud, $\mathcal{P}$ represents the GT point cloud, $\vert\mathcal{P}\vert$ signifies the number of points in $\mathcal{P}$, and $\Vert\cdot\Vert$ denotes L2 norm.
Similarly, the P2M Distance consists of two components: the average distance from each point in the predicted point cloud to the nearest face in the GT point cloud (referred to as P2F), and the sum of the average distances from each face in the GT point cloud to the nearest point in the predicted point cloud (referred to as F2P), as represented by Equation~\ref{eq4.2}.
\begin{equation}\label{eq4.2}
\begin{aligned}
{\mathcal{L}}_{P2M}(\hat{\mathcal{P}},\mathcal{M}) = {\mathcal{L}}_{P2F}(\hat{\mathcal{P}}&,\mathcal{M})+{\mathcal{L}}_{F2P}(\hat{\mathcal{P}},\mathcal{M}),\\
{\mathcal{L}}_{P2F} (\hat{\mathcal{P}},\mathcal{M})=\frac{1}{\vert\hat{\mathcal{P}}\vert}&\sum_{{\hat{p}\in\hat{\mathcal{P}}}} \min_{{f} \in \mathcal{M}} d(\hat{p},f),\\
{\mathcal{L}}_{F2P}(\hat{\mathcal{P}},\mathcal{M}) = \frac{1}{\vert\mathcal{M}\vert}&\sum_{{f} \in \mathcal{M}} \min_{{\hat{p}\in\hat{\mathcal{P}}}} d(\hat{p},f),
\end{aligned}
\end{equation}
where $\hat{\mathcal{P}}$ represents the predicted point cloud, $\mathcal{M}$ represents the mesh of GT point cloud, $\vert\mathcal{P}\vert$ signifies the number of points in $\mathcal{P}$, $\vert\mathcal{M}\vert$ signifies the number of faces in $\mathcal{M}$, and $d(p,f)$ denotes square distance from point $p$ to face $f$. Please note that due to the typically unequal quantities of faces and points, point-wise calculation of P2M distance is unreasonable. In the visualization of the denoising results of our experiments, we utilize the point-wise P2F distance from Equation~\ref{eq4.2} to visualize the denoised point cloud results.

\subsection{Comparison Methods}

Table \ref{compare_tech} shows the technical comparison with off-the-shelf methods.
\section{Further Comparisons}

\subsection{Additional Comparisons on Highly-dense and Large-scale Datasets}

As shown in Figure~\ref{fig:viewasiandragon}, Figure~\ref{fig:viewthailucy}, and Table~\ref{exp_stan_appendix}, additional experiments on Stanford 3D Scanning Repository are tested. 
In Figure~\ref{fig:viewasiandragon} and Figure~\ref{fig:viewthailucy}, local details are magnified to showcase the superior visual effects of 3DMambaIPF on filtering results around the surfaces of the point clouds.
Note that Pointfilter fails to process the large-scale datasets, Thai-Statue and Lucy.
While it achieved the best quantitative results on Asian-Dragon, its visual effect shows poor geometric details, as illustrated in Figure~\ref{fig:viewasiandragon}.
Significant noise reduction effects are not observed with PD-Flow and ScoreDenoise.
Although IterativePFN achieves competitive quantitative results, its visualization is marred by numerous surface holes. 
In contrast, 3DMambaIPF effectively resolves this issue.

\begin{table}[t]

\centering
\resizebox{\linewidth}{!}{
\begin{tabular}{c|cccccc}
\toprule [1.5pt]
\multirow{3}{*}{Method}

&\multicolumn{2}{c}{Asian-Dragon}
&\multicolumn{2}{c}{Lucy}
&\multicolumn{2}{c}{Thai-Statue}
  \\\cline{2-7}
  &\multicolumn{2}{c}{500K points}&\multicolumn{2}{c}{500K points}&\multicolumn{2}{c}{500K points}\\ 
\cline{2-7}
&CD&P2M&CD&P2M&CD&P2M \\ \toprule [1pt]
Noisy &50.19 &41.94&19.37&18.96&21.69&20.96\\
Pointfilter &\textbf{9.50}&\textbf{8.89}&-&-&-&-\\
PD-Flow &56.86&56.02&25.65&25.20&29.67&28.90\\
ScoreDenoise &41.49&40.75&18.57&18.17&21.63&20.92\\

IterativePFN &\underline{26.99}&\underline{26.29}&\underline{13.74}&\underline{13.35}&\underline{14.64}&\underline{14.01}\\\toprule [1pt]
3DMambaIPF (Ours) &27.66&26.95&\textbf{13.52}&\textbf{13.13}&\textbf{14.27}&\textbf{13.65}\\\toprule [1.5pt]
\end{tabular}
}
\caption{Additional Results on Stanford 3D Scanning Repository. CD and P2M distances are multiplied by $\textbf{10}^{\textbf{5}}$.}
\label{exp_stan_appendix}

\end{table}

\subsection{Visual Comparisons on Real-world Dataset}
Paris-rue-Madame dataset~\cite{serna2014paris}, comprising authentic street scenes captured in Paris via a 3D mobile laser scanner, is employed in the additional experiments.
As shown in Figure~\ref{fig:viewrue}, 3DMambaIPF and other baselines are tested on Paris-rue-Madame dataset.
Considering that the Paris-rue-Madame dataset is a real-world dataset captured by sensors and is inherently noisy, GT (noise-free) point clouds are not available for quantitative evaluation.
In the visualization of the results on Paris-rue-Madame, when there are no additional noises introduced, all the baseline methods are capable of achieving satisfactory filtering results.

\begin{table}[t]

\centering
\resizebox{\linewidth}{!}{
\begin{tabular}{c|c|cccccc}
\toprule [1.5pt]
\multirow{3}{*}{\makecell[c]{Rendering\\ Views}}
&\multirow{3}{*}{\makecell[c]{Mamba\\ Layers}}
&\multicolumn{6}{c}{50K points}\\ \cline{3-8}
&
&\multicolumn{2}{c}{1$\%$ noise}
&\multicolumn{2}{c}{2$\%$ noise}
&\multicolumn{2}{c}{2.5$\%$ noise}  \\ \cline{3-8}
&&CD&P2M&CD&P2M&CD&P2M \\ \toprule [1pt]

/ &/&20.56& 5.01& 30.43& 8.45& 33.52& 10.45\\\toprule [1pt]

/ &6&20.02&4.80&30.19&8.13&32.97&10.05\\

8& 6     &20.36&4.96 &30.73  &8.48 & 34.35 &10.96   \\

 16 &6  &20.29 &4.97 &30.35  &8.24 &33.08  &10.21 \\

24&6  &19.96  &4.82 &30.26  &8.14 &32.85  &9.95 \\

 32&6 &\textbf{19.89}& \textbf{4.77} &\textbf{29.95}&\textbf{8.03}   &\textbf{32.62}&\textbf{9.92}\\\midrule [1pt]

 32 & 1 & 23.35 & 6.76& 33.24 &10.91 &36.42   &13.17   \\

32& 3 &20.24 &5.02  &30.59 &8.61  &33.44 & 10.60  \\

32& 6 &\textbf{19.89}& \textbf{4.77} &\textbf{29.95}&\textbf{8.03}   &\textbf{32.62}&\textbf{9.92}\\

32 & 9 & 20.55&5.11 & 30.77 &8.59 &33.82  & 10.70   \\

32 & 12   &20.48&5.05 & 30.83 &8.80 &33.92  &10.98 \\\bottomrule [1.5pt]

\end{tabular}
}
\caption{Ablation results of Mamba layers and rendered views on PU-Net 10K dataset. CD and P2M distances are multiplied by $10^{5}$.}

\label{total_ablation_appendix}

\end{table}

\subsection{Detailed Results on PU-Net Dataset}
Results of all point clouds in the PU-Net dataset are detailly listed in Table~\ref{appendix_punet_mamba}. Comparisons with IterativePFN are shown in Table~\ref {ap_punet10} and Table~\ref{ap_punet50}. 3DMambaIPF demonstrated an overwhelming advantage over IterativePFN on the PU-Net dataset. 

\begin{table*}[t]

\centering
\resizebox{\linewidth}{!}{
\begin{tabular}{c|c|cccccc|cccccc}
\toprule [1.5pt]
\multirow{3}{*}{Method}&
\multirow{3}{*}{\makecell[c]{Patch\\ Size}}
&\multicolumn{6}{c|}{10K points}
&\multicolumn{6}{c}{50K points}\\ 
&&\multicolumn{2}{c}{1$\%$ noise}
&\multicolumn{2}{c}{2$\%$ noise}
&\multicolumn{2}{c|}{2.5$\%$ noise}
&\multicolumn{2}{c}{1$\%$ noise}
&\multicolumn{2}{c}{2$\%$ noise}
&\multicolumn{2}{c}{2.5$\%$ noise}  \\ 
&&CD&P2M&CD&P2M&CD&P2M&CD&P2M&CD&P2M&CD&P2M \\ \toprule [1pt]
IterativePFN&1000 &20.55& 5.01& 30.43& 8.43& 33.53& 10.46& 6.05 &3.02 &8.03&4.36 &10.15& 5.88\\
3DMambaIPF (Ours)&1000&\textbf{19.89}& \textbf{4.77} &\textbf{29.95}&\textbf{8.03}   &\textbf{32.62}&\textbf{9.92}& \textbf{5.89}&\textbf{2.91} &7.55&\textbf{4.05}  &9.28&5.31 \\\toprule [1pt]
IterativePFN&2000 &20.64&5.02&30.43&8.41&33.93&10.72&6.01&2.99&7.79&4.19&10.25&5.95 \\
3DMambaIPF (Ours)&2000& 19.98&4.81 &30.05&8.14 &32.73 &10.04& 5.92 &2.92 &\textbf{7.52}&\textbf{4.05}& \textbf{8.98}& \textbf{5.13}\\\toprule [1.5pt]

\end{tabular}
}
\caption{Ablation results of patch size on PU-Net Dataset. CD and P2M distances are multiplied by $\textbf{10}^{\textbf{5}}$.}
\label{ab_patch}

\end{table*}
\begin{table*}[h]

\centering

\begin{tabular}{c|cccccc|cccccc}

\toprule [1.5pt]

\multirow{3}{*}{Label}
&\multicolumn{6}{c|}{PU-Net 10K points}
&\multicolumn{6}{c}{PU-Net 50K points}\\
&\multicolumn{2}{c}{1$\%$ noise}
&\multicolumn{2}{c}{2$\%$ noise}
&\multicolumn{2}{c|}{2.5$\%$ noise}
&\multicolumn{2}{c}{1$\%$ noise}
&\multicolumn{2}{c}{2$\%$ noise}
&\multicolumn{2}{c}{2.5$\%$ noise}  \\ 
&CD&P2M&CD&P2M&CD&P2M&CD&P2M&CD&P2M&CD&P2M \\ \toprule [1pt]

Icosahedron   &25.24&3.37&44.98&6.00&48.75&7.90  &9.16&1.78&10.58&2.45&11.67&3.16\\
Octahedron    &22.40&4.00&32.94&5.09&34.81&6.01  &6.77&3.10&7.56&3.48&8.63&4.22\\
camel         &14.18&5.83&19.99&11.81&23.79&16.17   &3.48&2.02&6.00&4.46&9.22&7.06\\
casting       &25.57&5.52&40.88&12.07&45.33&15.03  &7.32&2.94&12.88&6.65&17.42&9.48\\
chair         &15.50&4.20&21.46&8.77&23.31&10.23     &5.13&2.68&6.60&3.86&10.32&6.52\\
coverrear\_Lp &22.80&3.37&35.27&7.32&37.94&9.79    &6.79&1.52&8.15&2.36&9.26&3.37\\
cow           &16.23&5.62&21.74&9.26&24.47&11.55     &4.12&3.09&5.82&4.43&7.72&5.92\\
duck          &22.78&6.62&34.12&8.69&35.70&9.80      &6.93&5.20&8.09&5.87&8.72&6.30\\
eight        &17.13&3.54&21.06&4.53&22.21&5.43     &4.38&2.72&5.12&3.16&5.93&3.71\\
elephant     &16.07&3.38&22.22&7.47&25.71&10.70     &4.00&1.50&5.82&2.94&8.61&5.02\\
elk           &22.98&5.63&35.13&9.86&37.01&10.79      &6.71&3.28&8.88&4.72&10.06&5.56\\
fandisk     &17.32&3.11&23.19&5.75&25.96&7.43            &4.35&1.80&5.44&2.49&7.32&3.83\\
genus3       &19.60&3.11&26.90&4.61&28.95&5.90        &5.35&2.16&6.22&2.72&7.23&3.40\\
horse       &13.21&3.36&17.58&6.76&19.13&8.02          &3.14&1.48&4.74&2.69&7.36&4.66\\
kitten        &20.21&3.92&27.92&7.06&30.29&9.11    &5.46&2.00&6.68&2.93&7.62&3.71\\
moai         &20.30&4.38&27.64&7.37&30.59&9.62    &5.56&2.40&7.22&3.58&8.20&4.26\\
pig           &18.83&7.70&25.53&11.01&27.10&12.19    &4.87&6.14&6.10&7.08&7.16&7.91\\
quadric       &16.74&5.27&22.32&10.07&25.81&13.03    &4.12&2.18&6.00&4.08&8.38&5.99\\
sculpt        &27.37&8.75&60.62&11.01&65.99&12.47    &12.65&6.72&14.66&7.34&15.77&8.02\\
star          &23.31&4.60&37.52&6.10&39.47&7.13       &7.58 &3.42&8.47&3.80&8.99&4.18\\ \toprule [1pt]
Mean          &19.89&4.77&29.95&8.03&32.62&9.92      &5.89&2.91&7.55&4.05&9.28&5.31\\

\toprule [1.5pt]

\end{tabular}
\caption{Detailed results on PU-Net Dataset. CD and P2M distances are multiplied by $\textbf{10}^{\textbf{5}}$.}

\label{appendix_punet_mamba}
\end{table*}
\begin{table*}[h]

\centering
\begin{tabular}{c|cccccc|cccccc}

\toprule [1.5pt]
\multirow{3}{*}{Label}
&\multicolumn{6}{c|}{IterativePFN}
&\multicolumn{6}{c}{3DMambaIPF (Ours)}\\ 
&\multicolumn{2}{c}{1$\%$ noise}
&\multicolumn{2}{c}{2$\%$ noise}
&\multicolumn{2}{c|}{2.5$\%$ noise}
&\multicolumn{2}{c}{1$\%$ noise}
&\multicolumn{2}{c}{2$\%$ noise}
&\multicolumn{2}{c}{2.5$\%$ noise}  \\ 
&CD&P2M&CD&P2M&CD&P2M&CD&P2M&CD&P2M&CD&P2M \\ \toprule [1pt]

Icosahedron   &26.57&3.49&44.92&6.05 &48.59& 7.88&25.24&3.37&44.98&6.00&48.75&7.90\\
Octahedron    &23.36&3.95&33.09&5.25 &36.22& 6.91&22.40&4.00&32.94&5.09&34.81&6.01\\
camel         &14.70&6.80&20.58&12.47&24.42&15.82&14.18&5.83&19.99&11.81&23.79&16.17\\
casting       &26.43&6.18&43.57&14.92&49.41&17.78&25.57&5.52&40.88&12.07&45.33&15.03\\
chair         &17.52&5.92&23.06&10.20&24.05&10.52&15.50&4.20&21.46&8.77&23.31&10.23\\
coverrear\_Lp &23.75&3.56&35.53&7.59 &38.70&10.25&22.80&3.37&35.27&7.32&37.94&9.79\\
cow           &16.50&5.95&21.90& 9.41&25.40&12.18&16.23&5.62&21.74&9.26&24.47&11.55\\
duck          &23.56&6.59&34.27& 8.75&36.38&10.15&22.78&6.62&34.12&8.69&35.70&9.80\\
eight         &17.29&3.52&21.49& 4.79&23.32& 6.25&17.13&3.54&21.06&4.53&22.21&5.43\\
elephant      &16.25&3.56&22.92& 8.11&26.61&11.05&16.07&3.38&22.22&7.47&25.71&10.70\\
elk           &23.91&6.04&35.60&10.37&38.11&11.75&22.98&5.63&35.13&9.86&37.01&10.79\\
fandisk       &17.60&3.16&23.49& 6.01&26.44&7.75&17.32&3.11&23.19&5.75&25.96&7.43\\
genus3        &20.15&3.09&27.06& 4.67&30.27&6.76&19.60&3.11&26.90&4.61&28.95&5.90\\
horse         &13.31&3.51&17.65& 6.77&19.65&8.24&13.21&3.36&17.58&6.76&19.13&8.02\\
kitten        &20.56&3.94&28.01& 7.11&30.76&9.43&20.21&3.92&27.92&7.06&30.29&9.11\\
moai          &20.67&4.44&27.81& 7.48&30.94&9.81&20.30&4.38&27.64&7.37&30.59&9.62\\
pig           &18.93&7.72&25.40&10.70&27.17&12.20&18.83&7.70&25.53&11.01&27.10&12.19\\
quadric       &16.91&5.38&22.61&10.39&26.60&13.58&16.74&5.27&22.32&10.07&25.81&13.03\\
sculpt        &28.65&8.74&61.78&11.28&67.18&13.05&27.37&8.75&60.62&11.01&65.99&12.47\\
star          &24.41&4.62&37.79& 6.36&40.32& 7.83&23.31&4.60&37.52&6.10&39.47&7.13\\ \toprule [1pt]
Mean          &20.55&5.01&30.43&8.43&33.53&10.46&19.89&4.77&29.95&8.03&32.62&9.92\\

\toprule [1.5pt]

\end{tabular}

\caption{Results comparing with IterativePFN on PU-Net 10K Dataset. CD and P2M distances are multiplied by $\textbf{10}^{\textbf{5}}$.}

\label{ap_punet10}
\end{table*}

\begin{table*}[h]

\centering
\begin{tabular}{c|cccccc|cccccc}

\toprule [1.5pt]
\multirow{3}{*}{Label}
&\multicolumn{6}{c|}{IterativePFN}
&\multicolumn{6}{c}{3DMambaIPF (Ours)}\\ 
&\multicolumn{2}{c}{1$\%$ noise}
&\multicolumn{2}{c}{2$\%$ noise}
&\multicolumn{2}{c|}{2.5$\%$ noise}
&\multicolumn{2}{c}{1$\%$ noise}
&\multicolumn{2}{c}{2$\%$ noise}
&\multicolumn{2}{c}{2.5$\%$ noise}  \\ 
&CD&P2M&CD&P2M&CD&P2M&CD&P2M&CD&P2M&CD&P2M \\ \toprule [1pt]

Icosahedron   &9.12&1.75&10.67&2.52&11.68&3.12&9.16&1.78&10.58&2.45&11.67&3.16\\
Octahedron    &6.71&3.05&7.75&3.58&9.07&4.48&6.77&3.10&7.56&3.48&8.63&4.22\\
camel         &3.56&2.13&6.52&4.86&10.73&8.10&3.48&2.02&6.00&4.46&9.22&7.06\\
casting       &7.62&3.27&15.18&8.11&19.75&10.67&7.32&2.94&12.88&6.65&17.42&9.48\\
chair         &8.09&4.66&7.61&4.33&10.78&6.80&5.13&2.68&6.60&3.86&10.32&6.52\\
coverrear\_Lp &6.82&1.52&8.68&2.70&10.13&3.89&6.79&1.52&8.15&2.36&9.26&3.37\\
cow           &4.13&3.16&5.92&4.53&8.49&6.53&4.12&3.09&5.82&4.43&7.72&5.92\\
duck          &6.89&5.18&8.19&5.93&9.17&6.59&6.93&5.20&8.09&5.87&8.72&6.30\\
eight         &4.33&2.68&5.31&3.28&6.44&4.02&4.38&2.72&5.12&3.16&5.93&3.71\\
elephant     &4.00&1.51&6.42&3.36&10.47&6.40&4.00&1.50&5.82&2.94&8.61&5.02\\
elk           &6.73&3.32&10.96&5.92&11.65&6.44&6.71&3.28&8.88&4.72&10.06&5.56\\
fandisk     &4.34&1.77&5.80&2.71&8.33&4.51&4.35&1.80&5.44&2.49&7.32&3.83\\
genus3       &5.33&2.16&6.46&2.87&7.98&3.91&5.35&2.16&6.22&2.72&7.23&3.40\\
horse       &3.14&1.49&4.91&2.81&8.70&5.70&3.14&1.48&4.74&2.69&7.36&4.66\\
kitten        &5.43&2.00&6.85&3.07&8.18&4.10&5.46&2.00&6.68&2.93&7.62&3.71\\
moai          &5.52&2.39&7.35&3.67&8.83&4.72&5.56&2.40&7.22&3.58&8.20&4.26\\
pig           &4.80&6.10&6.07&7.03&7.56&8.14&4.87&6.14&6.10&7.08&7.16&7.91\\
quadric       &4.14&2.25&6.35&4.46&9.49&6.83&4.12&2.18&6.00&4.08&8.38&5.99\\
sculpt        &12.70&6.69&14.97&7.52&16.29&8.34&12.65&6.72&14.66&7.34&15.77&8.02\\
star          &7.51&3.36&8.56&3.87&9.28&4.37&7.58&3.42&8.47&3.80&8.99&4.18\\ \toprule [1pt]
Mean          &6.05&3.02&8.03&4.36&10.15&5.88&5.89&2.91&7.55&4.05&9.28&5.31\\

\toprule [1.5pt]

\end{tabular}

\caption{Results comparing with IterativePFN on PU-Net 50K Dataset. CD and P2M distances are multiplied by $\textbf{10}^{\textbf{5}}$.}

\label{ap_punet50}
\end{table*}

\section{Additional Ablations}

\subsection{Ablations on PU-Net 10K dataset}
To verify the filtering capability of 3DMambaIPF on small-scale datasets, we also conducted ablation experiments on the PU-Net 10K dataset, as shown in Table~\ref{total_ablation_appendix}. Similar to the main paper, the impact of the number of Mamba layers and rendering views on the filtering capability is investigated.

\subsection{Ablations on Patch Size}
As shown in Table~\ref{ab_patch}, an additional ablation study on patch size is added. The experiments are tested on PU-Net 10K and 50K datasets with an NVIDIA GeForce RTX 3090 GPU. 
It is observed that a patch size of 2000 is more effective for filtering large-scale point clouds compared to a patch size of 1000.

\end{document}